\documentclass[]{jot}

\usepackage{multirow}
\usepackage[utf8]{inputenc}
\usepackage{acronym}
\usepackage{url}
\usepackage{color}
\usepackage{amssymb}
\usepackage{pifont}
\usepackage{graphicx}
\usepackage{floatflt} 
\usepackage{float}
\usepackage{framed}
\usepackage{etoolbox}
\usepackage{rotating}
\usepackage{booktabs}

\jotdetails{
    volume=19,      
    number=1,       
    articleno=e1,   
    year=2020,      
    license=ccby    
}

\definecolor{officegreen}{rgb}{0.0, 0.5, 0.0}
\definecolor{mordantred}{rgb}{0.68, 0.05, 0.0}

\newcommand{\cmark}{\ding{51}}%
\newcommand{\xmark}{\ding{55}}%
\newcommand{\hook}{\textcolor{officegreen}{\cmark}}
\newcommand{\cross}{\textcolor{mordantred}{\xmark}}

\articletype{manual} 

\newacro{MDE}{Model-Driven Engineering}
\newacro{SLR}{Systematic Literature Review}
\newacro{DBLP}{Digital Bibliography \& Library Project}
\newacro{CSV}{Comma-Separated Values}
\newacro{EMSL}{eMoflon Specification Language}
\newacro{PL}{Programming Languages}
\newacro{UML}{Unified Modeling Language}
\newacro{SE}{Software Engineering}
\newacro{IDE}{Integrated Development Environment}
\newacro{OCL}{Object Constraint Language}
\newacro{SMT}{Satisfiability Modulo Theories}
\newacro{CPS}{Cyber-Physical System}
\newacro{IFML}{Interaction Flow Modeling Language}
\newacro{SysML}{Systems Modeling Language}
\newacro{QVTo}{QVT Operational}
\newacro{SOA}{Service-Oriented Architecture}
\newacro{IoT}{Internet of Things}
\newacro{EPFC}{Eclipse Process Framework Composer}

\begin{document}
	
	\title{Tolerance in Model-Driven Engineering: A Systematic Literature Review with Model-Driven Tool Support}

\author[$\ast$]{Nils Weidmann}
\author[$\ast$]{Suganya Kannan}
\author[$\dagger$]{Anthony Anjorin}

\affil[$\ast$]{Paderborn University, Paderborn, Germany}
\affil[$\dagger$]{IAV GmbH Ingenieurgesellschaft Auto und Verkehr, Berlin, Germany}


\runningtitle{Tolerance in MDE: A Systematic Literature Review with Model-Driven Tool Support} 

\runningauthor{Weidmann \textit{et al.}}

\begin{abstract}
	Managing models in a consistent manner is an important task in the field of Model-Driven Engineering (MDE).
	Although restoring and maintaining consistency is desired in general, recent work has pointed out that always strictly enforcing consistency at any point of time is often not feasible in real-world scenarios, and sometimes even contrary to what a user expects from a trustworthy MDE tool.
	The challenge of tolerating inconsistencies has been discussed from different viewpoints within and outside the modelling community, but there exists no structured overview of existing and current work in this regard.
	In this paper, we provide such an overview to help join forces tackling the unresolved problems of tolerating inconsistencies in MDE.
	We follow the standard process of a Systematic Literature Review (SLR) to point out what tolerance means, how it relates to uncertainty, which examples for tolerant software systems have already been discussed, and which benefits and drawbacks tolerating inconsistencies entails.
	Furthermore, we propose a tool-chain that helps conducting SLRs in computer science and also eases the reproduction of results.
	Relevant meta-data of the collected sources is uniformly described in a textual modelling language and exported to the graph database Neo4j to query aggregated information.
\end{abstract}

%
%
	
	
%
%
%

	\maketitle
	
	\section{Introduction}
\label{sec:intro}

In the domain of \ac{MDE}, consistency management is an important challenge when multiple semantically interrelated models are to be developed and maintained simultaneously.
Consistency management involves multiple operations on models, such as (unidirectional) transformation, (concurrent) model synchronisation (propagating changes between models), and consistency checking.
Up until now, fundamental research has focussed on preserving or restoring perfect consistency between the involved models when performing such operations on them.
While eventually having consistent models in a software system is a desirable goal in general, several limitations reveal an 
apparent need for some form of fault-tolerance in practice.
In prior work, Stevens~\cite{BidirectionallyToleratingInconsistency-PartialTransformations.} argues for tolerating inconsistencies by listing several convincing scenarios.
In the simplest case, the underlying consistency relation is partial, which means it is not always possible for a model management tool to return a consistent pair of models.
In more complex cases, a consistent solution might exist, but the tool might not be able to return it, e.g. due to time restrictions.
There might be various consistent solutions, but it might be neither satisfactory to choose one of them at random, nor to present (potentially thousands of) equally good solutions to the user to choose one.
Closely related, but yet significantly different, is the concept of uncertainty in models~\cite{Partialmodels-Towardsmodelingandreasoningwithuncertainty.}.
In contrast to (temporarily) working with imperfect models, support for working with uncertainty allows a range of possible solutions to be encoded into a single model.
This can be useful as certain information might be unknown at design time or be about to change during the development process.
As approaches proposing support for tolerance and uncertainty tackle similar problems, i.e. increasing the practicability of \ac{MDE} techniques, it is difficult to draw the line between such approaches.

Although the topic has been addressed in many ways and from different viewpoints~\cite{ToleratingInconsistency., InstantconsistencycheckingfortheUML, BidirectionallyToleratingInconsistency-PartialTransformations.,  OntheQuestforFlexibleModelling.}, there does not yet exist an overview of research on tolerance in \ac{MDE}.
Such an overview would facilitate further research for several reasons:
First of all, it can collect and aggregate already achieved results and thereby unify definitions, establish common examples, and motivate research that has been left open.
Second, fault-tolerance is a problem with a much broader scope than software modelling.
Proposed definitions, examples, and (dis-)advantages might originate from other software engineering domains. 
In the database community, for example, there has been long-term research on maintaining and restoring consistency, and performing operations in the presence of errors~\cite{Decker2011, Decker2017}.
Consequently, to avoid reinventing the wheel, research from related fields should be taken into account, as long as it matches the problem domain and transferring results to \ac{MDE} is possible.
To achieve these goals, an overview of existing research should provide a proper definition of tolerance and delineate between the terms tolerance and uncertainty.
A collection of formal and practical frameworks should be identified that support handling of fault-tolerance in software modelling.
To motivate tolerant system behaviour, a range of plausible examples is essential, which can be used in future approaches to facilitate comparison to existing work.
Finally, a critical discussion of taking tolerance into account when modelling software systems should also be included to identify opportunities and risks that are inherent to tolerant approaches.

To address these requirements, we present the results of a \ac{SLR} on tolerance in \ac{MDE} conducted from October 2019 to September 2020 following the standard methodology proposed by Kitchenham et al.~\cite{Kitchenham2004, Systematicliteraturereviewsinsoftwareengineering-Asystematicliteraturereview}.
With this review, we particularly aim at answering the following research questions:

\begin{itemize}
	\item [RQ1] \textbf{Scope and Classification}: 
	How is (in)consistency defined? 
	How do tolerance and uncertainty differ? 
	What makes an approach or tool tolerant? 
	Which different dimensions of tolerance are there and how can tolerance be classified?
	
	\item [RQ2] \textbf{Examples and Application Domains}: 
	In which application domains is tolerance relevant? 
	Which examples are used to demonstrate ideas and results?
	
	\item [RQ3] \textbf{Benefits and Challenges}: 
	What are the benefits of tolerance?
	What are open questions and challenges?
\end{itemize}

As a second contribution, we propose a framework that supported us while conducting this \ac{SLR} and which can be reused for creating future literature reviews in computer science.
In consensus with previous findings~\cite{Goetz2018}, we noticed that \acp{SLR} in general are conducted with little or no tool support (or at least lack a respective description), although they involve numerous steps that could be automated, leading to unnecessary manual effort.
Likewise, it also requires a substantial amount of work to reproduce the results of an \ac{SLR}, leading to opacity of findings due to time restrictions.
The gathering and reproduction of results should thus be eased to better utilise human resources for tasks that require advanced knowledge of the problem domain.
We propose a tool-chain for partly automating the review process, which involves an adapter for querying the research database, a transformation of the results to a modelling language processable by eMoflon::Neo\footnote{\url{https://github.com/eMoflon/emoflon-neo}}, a model management tool that is a recent addition to the eMoflon tool suite~\cite{Weidmann2019}, and an export to the graph database Neo4j\footnote{\url{https://neo4j.com/}}, which can be queried to analyse the results.

The remainder of this paper is structured as follows:
Section~\ref{sec:survey-procedure} describes the survey procedure and sketches the used tooling.
Summaries of the answers to the research questions are provided in Sect.~\ref{sec:definitions}, \ref{sec:examples} and \ref{sec:benefits-drawbacks} (a more detailed tabular overview is available online\footnote{\url{https://drive.google.com/file/d/1uSuOn3hX5BHpLhw3jaH2ZpVffgTxHOov}}).
Section~\ref{sec:result-analysis} briefly analyses meta-data of the included sources and motivates further research on tolerance in \ac{MDE}.
Section~\ref{sec:related-work} gives an overview of related work, before Sect.~\ref{sec:conclusion-future-work} concludes the paper.
	\section{Survey Procedure}
\label{sec:survey-procedure}

This section briefly presents the methodology we followed to conduct the \ac{SLR}.
As all results should be reproducible and easily accessible for researchers of the modelling community, we therefore followed the guidelines proposed by Kitchenham et al.~\cite{Systematicliteraturereviewsinsoftwareengineering-Asystematicliteraturereview} for literature reviews in the software engineering domain.
The review was conducted from October 2019 to September 2020 and considers sources published until June 2020.
We used DBLP as a research database due to its large amount of listed publications, its focus on computer science, and its well-described API for automated queries.\footnote{\url{https://dblp.uni-trier.de/faq/13501473.html}}
Following the proposed guidelines, an initial and a final set of sources was determined by applying search phrases, and criteria for inclusion, exclusion, and quality of the gathered sources, described in the following.

To form an \textbf{initial set of sources}, we defined six search strings inspired by the research questions and the domain \ac{MDE}, of which at least two must appear in a title.
Each of these strings has a wildcard (\texttt{*}) as suffix to take nouns, verbs and adjectives into account.
We therefore decided to query the database with all pairs formed from the search strings \texttt{model*}, \texttt{consisten*}, \texttt{inconsisten*}, \texttt{uncertain*}, \texttt{tolera*} and \texttt{flexib*}.
As a combined inclusion and quality criterion, we further require the respective sources to be published at a conference listed by the CORE ranking\footnote{\url{http://portal.core.edu.au/conf-ranks/}} and assigned to the research field 0803 (Software Engineering).
Due to this requirement, the search is focussed on the software engineering domain and peer-reviewed publications, while journal and workshop papers are initially excluded.
Due to the publishing behaviour in the computer science domain, we expect late-breaking research results to be published at conferences, whereas journal articles usually extend previously published results of conference papers in more depth.
Furthermore, the list of journals at CORE\footnote{\url{http://portal.core.edu.au/jnl-ranks/}} was outdated when the review was conducted, making it difficult to apply equal criteria to journal and conference papers in the initial search step.
Workshop papers often present work in progress and initial ideas to be published at conferences afterwards.
To detect relevant papers which do not fulfil all criteria of the initial search, we applied snowballing at a subsequent step.
In this manner, we retrieved 268 sources, which we denote in the following as \textit{core} papers.

To compile a \textbf{final set of sources}, we distributed the core papers between three researchers and assessed their relevance based on the abstract. 
In case it remained unclear if the respective paper should be considered, introduction and conclusion were read as well.
As suitable examples demonstrating the use of tolerant system behaviour are essential to answer RQ2, the remaining parts of the papers were skimmed for such examples.
The assessment was based on inclusion, exclusion, and quality criteria. 
A paper was included in the further review process if it (1) presents an \ac{MDE} or \ac{PL} approach related to (in)consistency management, or (2) if it contains an example or application related to tolerance or uncertainty.
We excluded a paper if any of the search terms has a different meaning than the one implied by the research questions, e.g. if model refers to the physical behaviour of a \ac{CPS}.
Papers written in other languages than English were also excluded, while this criterion never had to be applied, probably due to the choice of search strings.
To ensure a high quality of the selected sources, prefaces and extended abstracts were excluded.
In total, 114 \emph{relevant core papers} were identified and added to the final set of sources.

In a second iteration, we applied \textbf{snowballing} to consider papers that were not detected in our first iteration but might be nonetheless relevant to answer the research questions.
The corpus of this \ac{SLR} was extended by all sources which are cited by at least one of the core papers, resulting in 3201 additional papers.
To keep the number of paper for the second assessment phase manageable, we added a further inclusion criterion for these additional papers:
a minimum citation count by relevant core papers, as papers cited more frequently are more likely to be relevant.
As newer sources naturally have a lower citation count, the number of required citations was set in relation to the publication year.
As papers published at \ac{SE} venues should be preferred, the minimum citation count per year was set to 0.2 for \ac{SE} papers and to 0.3 for all other papers.
As a result, 53 papers from SE venues and 41 papers from other venues were evaluated according to the same inclusion, exclusion, and quality criteria as the core papers. 
After the second assessment phase, 23 papers from \ac{SE} venues and 20 other sources were added, increasing the final set of sources to 157 papers.
An overview of our assessment and selection process is depicted in Tab.~\ref{tab:categorization}.
For each property, a check (\hook) means that it is fulfilled by the respective category of papers, whereas a cross (\cross) means the opposite.
If the property is irrelevant, this is indicated by a hyphen (-).
The rightmost two columns contain the number of papers per category identified as (not) relevant.

\begin{table}[htb]
	\begin{scriptsize}
		\begin{tabular}{|c|c|c||c||c|c|}
			\hline
			$\geq 2$ keywords & published at  & min. annual    & initially & rele- & not re- \\
			in title?         & an SE venue?  & citation count & read?     & vant  & levant  \\
			\hline
			\hook  & \hook  & -          & \hook  & 114 &  154 \\
			\cross & \hook  & $\geq 0.2$ & \hook  &  23 &   30 \\
			\cross & \hook  & $< 0.2$    & \cross &   0 & 1093 \\
			-      & \cross & $\geq 0.3$ & \hook  &  20 &   21 \\
			-      & \cross & $< 0.3$    & \cross &   0 & 2014 \\
			\hline
		\end{tabular}
	\end{scriptsize}
	\caption{Categorization of papers for the review}
	\label{tab:categorization}
\end{table}

Although the survey procedure is well-defined and takes multiple objective measures into account in order to make results transparent and reproducible, several \textbf{threats to validity} have to be mentioned as well.
Firstly, the naming conventions for conferences in the DBLP database and in the CORE ranking differ slightly.
To overcome this problem, we matched the respective conferences if their acronym is the same, or if one name is a substring of the other.
This method works well for venues listed in the latest version of the CORE ranking but, as conferences are renamed over time, older venues might not be identified as \ac{SE} venues.
While prefaces and extended abstracts were excluded, no distinction was made between different paper categories, such as full, short or tool paper.
Even though late-breaking results are usually published as conference papers in the computer science domain, we might have missed results only published in journal papers, as those were not considered in the first assessment phase.
Besides DBLP, the use of other research databases - such as Google scholar - could have helped to gather more sources for the \ac{SLR} and to therefore minimise the risk of missing important work.
Finally, although inclusion, exclusion and quality criteria were discussed between the involved researchers in detail before the review was conducted, only one researcher per paper evaluated its relevance, which might lead to biased results as the assessment of relevance depends on a single person.

\begin{figure}[htb]
	\centering
	\includegraphics[width=0.5\columnwidth]{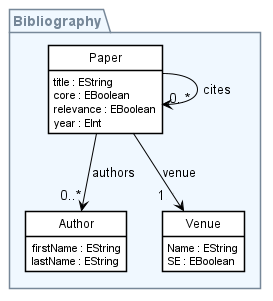}
	\caption{\ac{SLR} Metamodel}
	\label{fig:metamodel}
\end{figure}

\begin{figure}[htb]
	\centering
	\includegraphics[width=0.9\columnwidth]{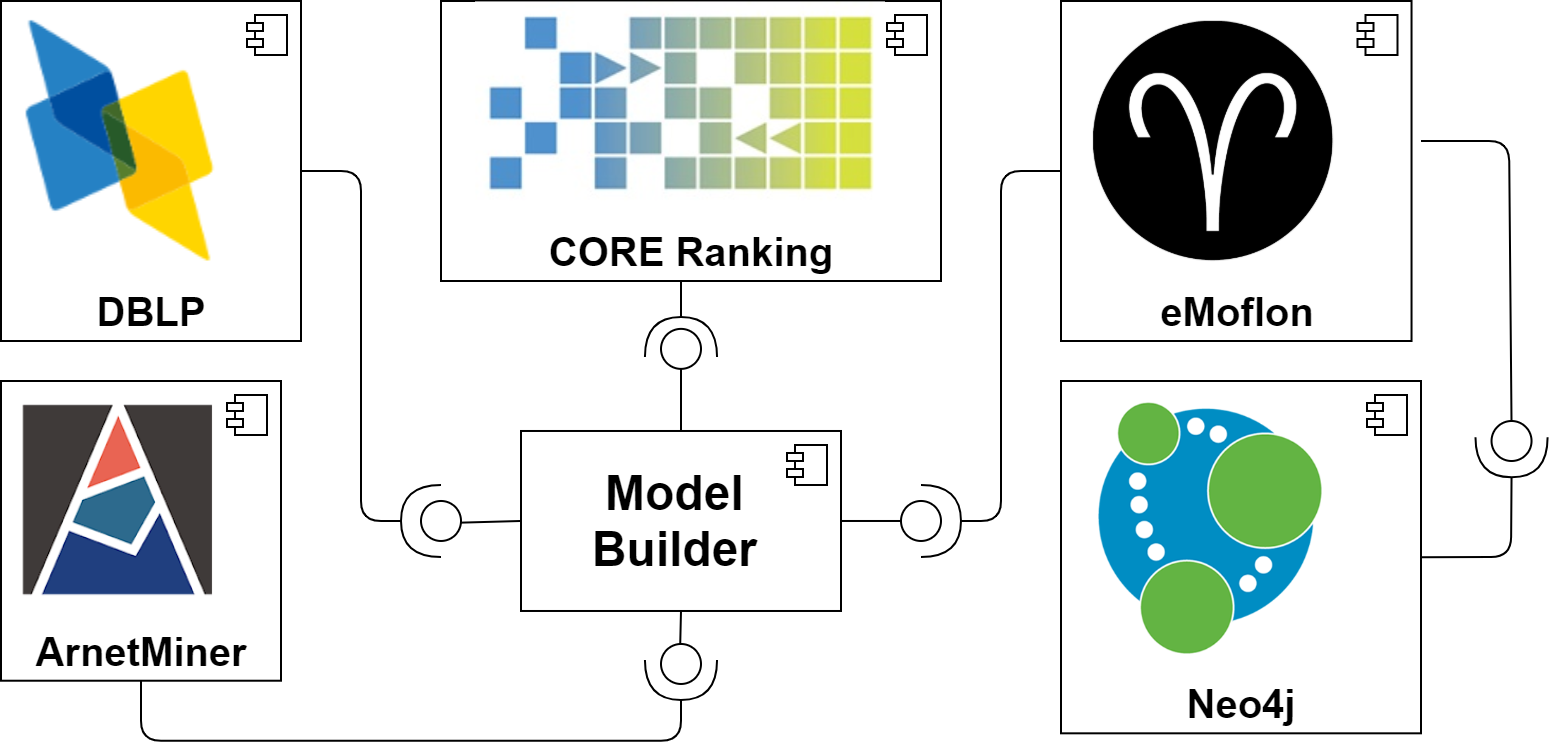}
	\caption{Component Diagram: Tool-chain}
	\label{fig:toolchain}
\end{figure}

In Fig.~\ref{fig:metamodel}, the metamodel for the representation of results is depicted.
A \texttt{Paper} is written by \texttt{Aut\-hor}s and appears at a \texttt{Venue}.
For each \texttt{Paper}, the title and the year of publication are extracted.
Boolean values indicate whet\-her the paper is a core paper, and whether it was identified as relevant for the \ac{SLR} in the initial reading phase.
A \texttt{cites} relation defines which \texttt{Paper} references which other \texttt{Paper}s.
For the \texttt{Author}, only the name is stored, whereas the \texttt{Venue} is additionally flagged as being an \ac{SE} conference listed at DBLP or not.
In Fig.~\ref{fig:toolchain}, the architecture for transferring the bibliography data records to a graph database is depicted.
The DBLP research database provides an interface to query records that match a specified search string containing logical connectors and wild cards.
Additionally, we exported the list of venues assigned to the research field 0803 (\ac{SE}) as a \ac{CSV} file.
For each possible pair of the six keywords, the database was queried and its venue was compared to the list of \ac{SE} venues.
In this way, the core papers for this \ac{SLR} are identified and saved as models typed over the metamodel of Fig.~\ref{fig:metamodel}.
After examining each of the core papers regarding its relevance for the research questions, the respective attribute was manually added in the bibliography model.
In the last step, the files were exported from eMoflon::Neo to Neo4j (cf. Sect.~\ref{sec:intro}), such that queries on the bibliography model can be used to analyse the collected data.

An important argument for the use of a database representation for the \ac{SLR} was the snowballing step, which is - depending on the number of relevant papers - a time-consuming and error-prone task.
To integrate citation information into the database, all papers referenced by the core papers were added via the database snapshot \textit{DBLP-Citation-network V12}\footnote{\url{https://www.aminer.org/citation}} which was created with the tool ArnetMiner~\cite{ArnetMiner:ExtractionAndMiningOfAcademicSocialNetworks}.
Having extended the bibliography model with all papers cited from the initial set of sources, it is possible to collect all papers that fulfil the condition for being added within the snowballing step with a single database query.
	\section{Scope and Classification}
\label{sec:definitions}

\begin{figure*}[htb]
	\begin{minipage}{0.8\textwidth}
		\includegraphics[width=\columnwidth]{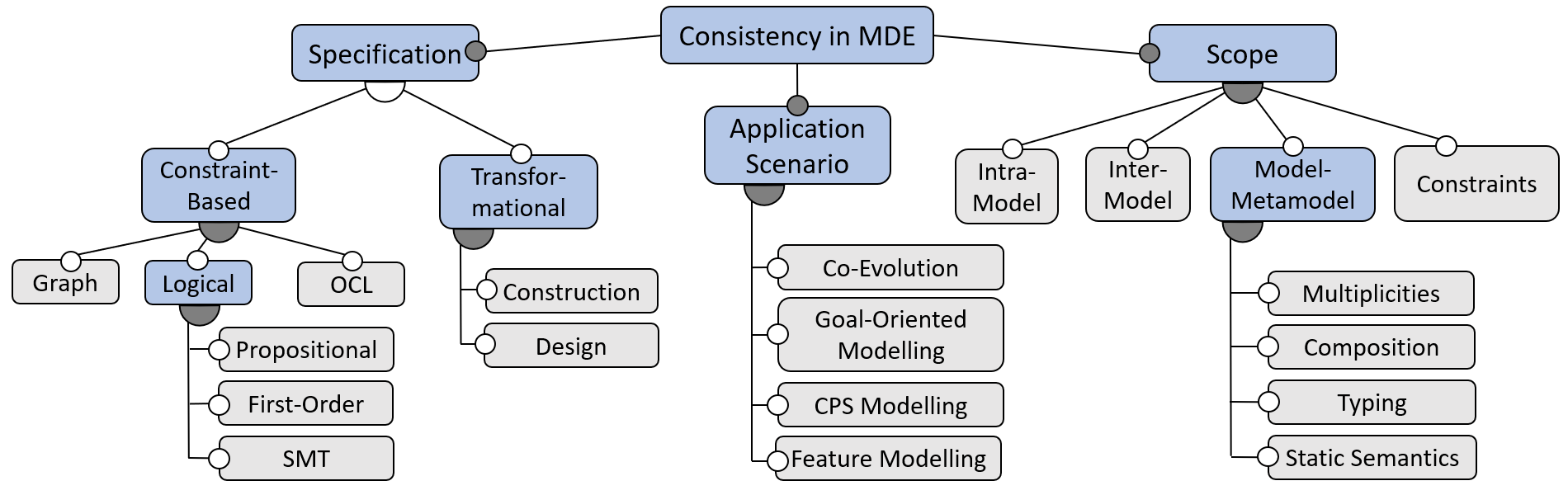}
	\end{minipage}
	\hfill
	\begin{minipage}{0.18\textwidth}
		\includegraphics[width=\columnwidth]{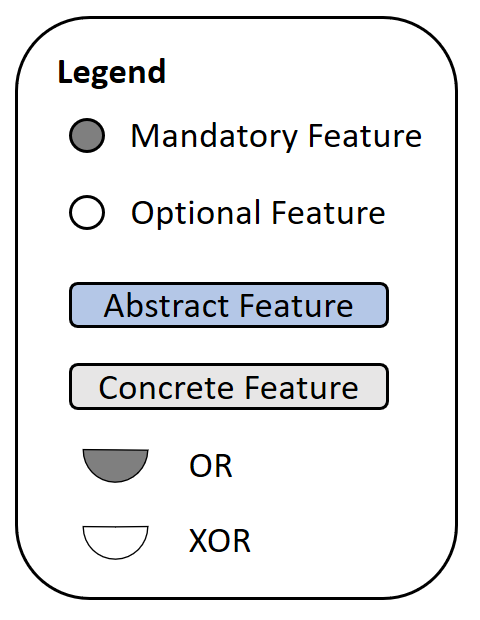}
	\end{minipage}
	\caption{Consistency in \ac{MDE}}
	\label{fig:feature-model-consistency}
	\vspace{-2mm}
\end{figure*}

\begin{figure*}[htb]
	\begin{minipage}{0.8\textwidth}
		\includegraphics[width=\columnwidth]{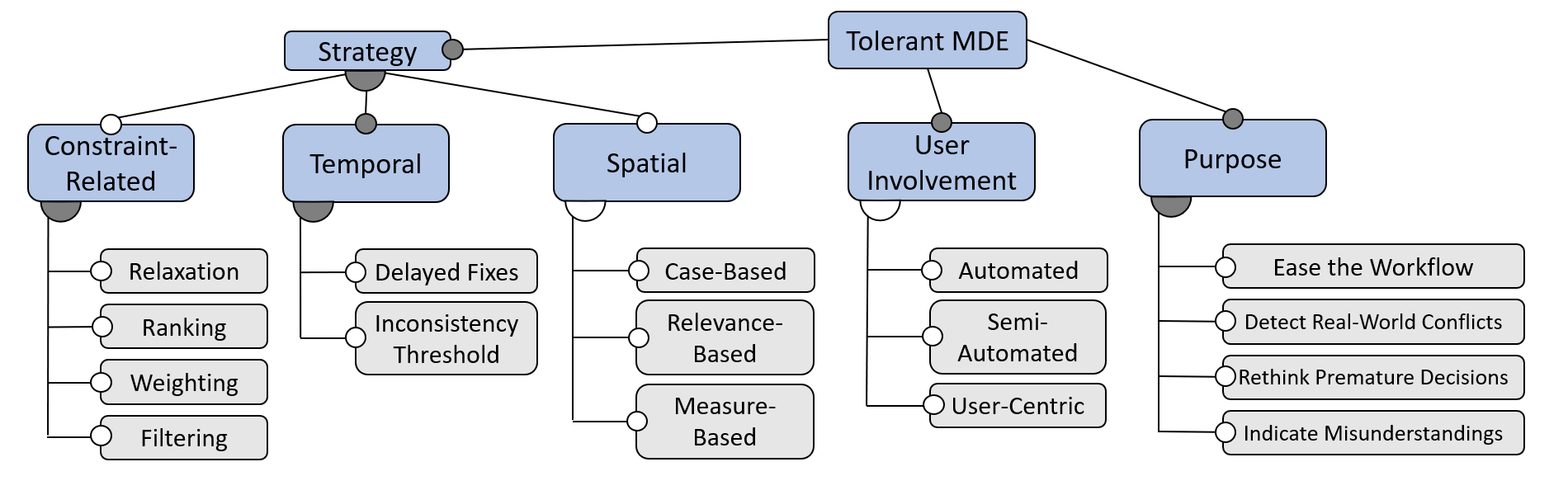}
	\end{minipage}
	\hfill
	\begin{minipage}{0.18\textwidth}
		\includegraphics[width=\columnwidth]{figures/feature-model-legend}
	\end{minipage}
	\caption{Tolerant \ac{MDE}}
	\label{fig:feature-model-tolerance}
	\vspace{-2mm}
\end{figure*}

\begin{figure*}[htb]
	\begin{minipage}{0.8\textwidth}
		\includegraphics[width=\columnwidth]{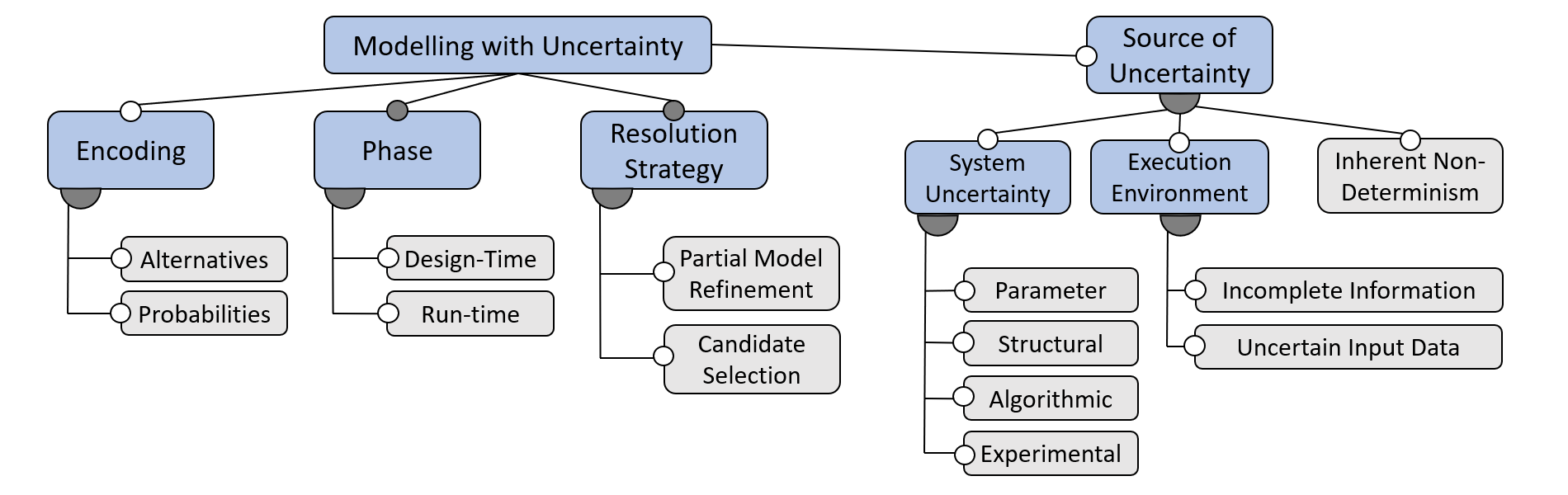}
	\end{minipage}
	\hfill
	\begin{minipage}{0.18\textwidth}
		\includegraphics[width=\columnwidth]{figures/feature-model-legend}
	\end{minipage}
	\caption{Modelling with Uncertainty}
	\label{fig:feature-model-uncertainty}
	\vspace{-2mm}
\end{figure*}

In order to answer the first research question, definitions for the key terms of this \ac{SLR} were gathered and aggregated while analysing the relevant sources.
Besides a brief summary, feature models for consistency (Fig.~\ref{fig:feature-model-consistency}), tolerance (Fig.~\ref{fig:feature-model-tolerance}) and uncertainty (Fig.~\ref{fig:feature-model-uncertainty}) were created to provide an overview of the different dimensions involved.

\subsection{Consistency}
In general, consistency can be understood as a relation over sets of models, which can be specified in different ways~\cite{BidirectionallyToleratingInconsistency-PartialTransformations., GlobalconsistencycheckingofdistributedmodelswithTReMer+., HeterogeneousModelsMatchingforConsistencyManagement.,Towardssoundoptimalandflexiblebuildingfrommegamodels, BidirectionalTransformationsintheLarge, IsBidirectionalityImportant,JTLabidirectionalandchangepropagatingtransformationlanguage}.
Most frequently, consistency is specified by a provided set of constraints~\cite{Anautomatedandinstantdiscoveryofconcreterepairsformodelinconsistencies., ApplyingaConsistencyCheckingFrameworkforHeterogeneousModelsandArtifactsinIndustrialProductLines., Computingrepairtreesforresolvinginconsistenciesindesignmodels., ConsistencyCheckinModellingMulti-AgentSystems., ConsistencyCheckingofConceptualModelsviaModelMerging., Constraint-BasedConsistencyCheckingbetweenDesignDecisionsandComponentModelsforSupportingSoftwareArchitectureEvolution., DataQualityMaintenancebyIntegrity-PreservingRepairsthatTolerateInconsistency., DetectingandResolvingModelInconsistenciesUsingTransformationDependencyAnalysis., Detectingmodelinconsistencythroughoperation-basedmodelconstruction., FineTuningModelTransformationChangePropagationinContextofConsistencyCompletenessandHumanGuidance., FixingInconsistenciesinUMLDesignModels., FromAbstracttoConcreteRepairsofModelInconsistencies-AnAutomatedApproach., GeneratingandEvaluatingChoicesforFixingInconsistenciesinUMLDesignModels., Inconsistencymanagementframeworkformodel-baseddevelopment., Inter-modelConsistencyCheckingUsingTripleGraphGrammarsandLinearOptimizationTechniques., Loosely-coupledConsistencybetweenAgent-orientedConceptualModelsandZSpecifications., ModelingandEnactmentSupportforEarlyDetectionofInconsistenciesinEngineeringProcesses., OntheQuestforFlexibleModelling., RuntimeAdjustmentofConfigurationModelsforConsistencyPreservation., SpecifyingConsistencyConstraintsforModellingLanguages., SupportingConsistencybetweenArchitecturalDesignDecisionsandComponentModelsthroughReusableArchitecturalKnowledgeTransformations., ToleratingInconsistency., TowardsModel-and-CodeConsistencyChecking., AutomaticallyDetectingandTrackingInconsistenciesinSoftwareDesignModels, Consistencymanagementwithrepairactions, Constraintprogrammingfortypeinferenceinflexiblemodeldrivenengineering, InstantconsistencycheckingfortheUML, Makinginconsistencyrespectableinsoftwaredevelopment, Supportingautomaticmodelinconsistencyfixing, xlinkitaconsistencycheckingandsmartlinkgenerationservice,AutomaticGenerationofAtomicConsistencyPreservingSearchOperatorsforSearch-BasedModelEngineering,AutomatedGenerationofConsistentGraphModelswithFirst-OrderLogicTheoremProvers, Detectingandexploringsideeffectswhenrepairingmodelinconsistencies,CheckingUMLandOCLModelConsistency-AnExperienceReportonaMiddle-SizedCaseStudy,QuickfixgenerationforDSMLs,DeterminingtheCauseofaDesignModelInconsistency}, which can be formulated in different ways.
Often, the \ac{OCL} is chosen for defining consistency, but also graph constraints and logical constraints, such as formulae of propositional or first-order logic or as \ac{SMT}, are commonly used.
Independent of the language in use, a model (or a proposed solution) is typically viewed as being consistent if it satisfies all constraints.

Consistency can also be defined constructively via a given model transformation $T$, such that two models $A$ and $B$ are consistent if and only if $A = T(B)$~\cite{AnIncrementalAlgorithmforHigh-PerformanceRuntimeModelConsistency., ImplementingQVTRbidirectionalmodeltransformationsusingalloy}.
Furthermore, multiple model transformations (e.g., syntactic changes) are often considered to be consistent if they implement the same underlying transformation (e.g., semantic change)~\cite{Consistentco-evolutionofmodelsandtransformations.}.
This is especially relevant in the context of co-evolution, where multiple interrelated transformations are concurrently conducted.
There are also constructive definitions which define methodological consistency over the sequence of operations required to construct a model.
As long as certain steps are followed in the construction or ``design'' process, the consistency of the resulting model(s) can be guaranteed~\cite{Detectingmodelinconsistencythroughoperation-basedmodelconstruction., IncrementalConsistencyCheckingforComplexDesignRulesandLargerModelChanges.}. 
For example, a name must be assigned immediately after an element is created.

Besides these general consistency specifications, some definitions are also tailored to a specific application area.
When modelling \acp{CPS}, a model is said to be consistent with the real world - or any other system for which this can be checked - if the model makes statements or reaches conclusions that are actually true~\cite{Acamerauncertaintymodelforcollaborativevisualsensornetworkapplications}.
For goal-oriented modelling, ``plan consistency'' means that the achievement of sub-goals implies the achievement of their parent goal~\cite{Declarativeprojectplanningandcontrolling-aformalmodeltosupportthehandlingofunavoidableinconsistencies.}.
In the application domain of software product line engineering, a feature model is consistent if at least one valid configuration exists~\cite{FlexibleModelingandProductDerivationinSoftwareProductLines.}. 


The notions of consistency can be applied to both the intra- and inter-model case.
For inter-model consistency, multiple models are consistent if they are not in conflict regarding their overlapping parts, i.e., the same information contained in multiple models~\cite{AnApproachforDetectingInconsistenciesbetweenBehavioralModelsoftheSoftwareArchitectureandtheCode.,ConsistentIntegrationofModelsBasedonViewsofVisualLanguages.,ConsistentModelingUsingMultipleUMLProfiles.,EvaluatingtheImpactofAspectsonInconsistencyDetectionEffort-AControlledExperiment.}. 
Another important definition deals with the relationship between a model and its meta-model~\cite{ComposingModelsforDetectingInconsistencies-ARequirementsEngineeringPerspective.,EnhancedAutomationforManagingModelandMetamodelInconsistency.,FlexibleModelElementIntroductionPoliciesforAspect-OrientedModeling.,Formalizingmodelconsistencybasedontheabstractsyntax.,ImprovingInconsistencyResolutionwithSide-EffectEvaluationandCosts.,OntheQuestforFlexibleModelling.,PrototypinganInconsistencyCheckingToolforSoftwareProcessModels.,TheConformanceRelationChallenge-BuildingFlexibleModellingFrameworks.,AMetamodelingFrameworkforPromotingFlexibilityandCreativityOverStrictModelConformance,CAREaconstraintbasedapproachforreestablishingconformancerelationships,IntroducingTraceabilityandConsistencyCheckingforChangeImpactAnalysisacrossEngineeringToolsinanAutomationSolutionCompanyAnExperienceReport,JSMFaJavascriptFlexibleModellingFramework,UsingDescriptionLogictoMaintainConsistencybetweenUMLModels,AutomaticGenerationofAtomicConsistencyPreservingSearchOperatorsforSearch-BasedModelEngineering,AutomatedGenerationofConsistentGraphModelswithFirst-OrderLogicTheoremProvers, Detectingandexploringsideeffectswhenrepairingmodelinconsistencies,CheckingUMLandOCLModelConsistency-AnExperienceReportonaMiddle-SizedCaseStudy,Asatisficingbidirectionalmodeltransformationengineusingmixedintegerlinearprogramming,DeterminingtheCauseofaDesignModelInconsistency,QuickfixgenerationforDSMLs}.
This notion can be handled on two levels:
(i) structural consistency includes multiplicities, composition constraints, as well as the types of model elements.
(ii) static semantics expressed, e.g. via \ac{OCL} constraints.
Finally, a set of constraints is often denoted as consistent if there exists at least one solution (e.g., a variable assignment) that satisfies all constraints~\cite{ADynamic-PriorityBasedApproachtoFixingInconsistentFeatureModels.,CheckingModelConsistencyUsingData-FlowTesting.,ConsistencyIndependenceandConsequencesinUMLandOCLModels,PrototypinganInconsistencyCheckingToolforSoftwareProcessModels.,UsingDescriptionLogictoMaintainConsistencybetweenUMLModels}.

\subsection{Tolerance}
Based on the collected definitions of consistency, the term tolerance can be specified more concretely.

Building on the constraint-based definition of consistency, tolerance can be implemented by weakening the requirement of constraint satisfaction~\cite{ADynamic-PriorityBasedApproachtoFixingInconsistentFeatureModels., ComposingModelsforDetectingInconsistencies-ARequirementsEngineeringPerspective., FlexibleModelingandProductDerivationinSoftwareProductLines., FlexibleUpdatePropagationforWeaklyConsistentReplication., IncrementalConsistencyCheckingforComplexDesignRulesandLargerModelChanges., Inter-modelConsistencyCheckingUsingTripleGraphGrammarsandLinearOptimizationTechniques., ModelingandEnactmentSupportforEarlyDetectionofInconsistenciesinEngineeringProcesses., Modelingexecutiontimeofmulti-stageN-versionfault-tolerantsoftware.}.
Solutions that satisfy more important constraints are then ``better'', i.e., more consistent than other solutions that might satisfy more but less important constraints.
Tolerance here is, therefore, basically a ranking or weighting of constraints and a score for solutions based on how many weighted constraints are satisfied.
This prioritisation and sorting process is often referred to as ``relaxation''; the constraints are sometimes denoted as ``soft constraints'' as their violation no longer directly implies exclusion of the respective solution.
By defining different classes of inconsistencies and measuring consistency as vectors over these dimensions, one can obtain a more fine-grained view of the extent to which consistency is achieved or improved with respect to each class \cite{Designandevaluationofacontinuousconsistencymodelforreplicatedservices, DetectingandRepairingInconsistenciesacrossHeterogeneousModels.}.
A similar approach divides the constraints or requirements into primary and non-primary, whereby non-primary constraints can be ignored in a tolerant scenario~\cite{AConsistencyModelforIdentityInformationinDistributedSystems., Inconsistencymanagementframeworkformodel-baseddevelopment.}.
The idea is to ignore inconsistencies that are ``irrelevant'', e.g., concerning white space or time stamps, layout, etc.

Another important group of strategies for implementing tolerance involve temporal aspects.
Most approaches assume that inconsistencies can be tolerated up to some point in time when consistency is restored, such that fixes are delayed up to this point~\cite{AConsistencyModelforIdentityInformationinDistributedSystems.,Designandevaluationofacontinuousconsistencymodelforreplicatedservices}.
For distributed systems, a variable threshold for inconsistencies is defined by a temporal window such that more inconsistencies are accepted at the beginning and fewer towards the end. 
These approaches aim at letting a system ``stabilise'' before demanding a high level of consistency.

In contrast to temporal strategies, a further group takes a spatial approach to implementing tolerance: case-based restoration guarantees that every part of the model that was consistent before is still consistent afterwards~\cite{BidirectionallyToleratingInconsistency-PartialTransformations., Classifyingintegritycheckingmethodswithregardtoinconsistencytolerance., DataQualityMaintenancebyIntegrity-PreservingRepairsthatTolerateInconsistency., BidirectionalTransformationsintheLarge, InstantconsistencycheckingfortheUML, ToleratingInconsistency.}.
For efficiency reasons, only a subset of ``relevant'' cases can be determined, i.e., a scope of influence is computed for changes, and then checked as for case-based restorers.
Measure-based restorers guarantee that a chosen measure of consistency is not reduced by the restoration process.

A frequently named property of fault-tolerant systems is that strategies are implemented to detect and fix inconsistencies either automatically or by interacting with the user~\cite{AFormalModelforFault-ToleranceinDistributedSystems.,GeneratingandEvaluatingChoicesforFixingInconsistenciesinUMLDesignModels.,OntheQuestforFlexibleModelling., OntheCollaborativeDevelopmentofPara-ConsistentConceptualModels.,Pattern-BasedModelingandAnalysisofFailsafeFault-ToleranceinUML.,UML/Analyzer-AToolfortheInstantConsistencyCheckingofUMLModels., xlinkitaconsistencycheckingandsmartlinkgenerationservice,AutomaticallyDetectingandTrackingInconsistenciesinSoftwareDesignModels,Modelingexecutiontimeofmulti-stageN-versionfault-tolerantsoftware.}.
Even if a system is brought into an inconsistent state, it can transition back to a consistent state by applying fix strategies.
Consequently, an inconsistent state is temporarily acceptable, making it unnecessary to check if edits are consistency-preserving, or to propagate changes to other models immediately.
However, it is often important to keep track of inconsistent model parts as this can speed up the consistency restoration later.
In case of user involvement, this can also help to avoid overwhelming users with too many design decisions.

Overall, tolerant concepts can help ease the work flow for modelling tasks by not enforcing an immediate resolution of inconsistencies. 
Additionally, they can function as a mechanism to detect unresolved conflicts in the real world, or to rethink prematurely made design decisions~\cite{Makinginconsistencyrespectableinsoftwaredevelopment}.
Finally, tolerating and highlighting inconsistencies can be used to indicate misunderstandings or a potential disagreement of the involved developers~\cite{Supportingautomaticmodelinconsistencyfixing}.

\subsection{Uncertainty}
To help make a distinction between tolerance and uncertainty, we provide an overview of the most common notions of uncertainty in the following.

Modelling with uncertainty often involves encoding a range of possible values or alternatives into a single attribute value or part of a model~\cite{DataUncertaintyModelforMashup., EnhancingFlexibilityinUserInteractionModelingbyAddingDesignUncertaintytoIFML., HandlingUncertaintyinAutomaticallyGeneratedImplementationModelsintheAutomotiveDomain, Mu-Mmint:anIDEformodeluncertainty, Managinguncertaintyinbidirectionalmodeltransformations, MAV-Vis:anotationformodeluncertainty, OntheCollaborativeDevelopmentofPara-ConsistentConceptualModels., Partialmodels-Towardsmodelingandreasoningwithuncertainty., TestModelCoverageAnalysisUnderUncertainty., The'domainmodelconcealer'and'applicationmoderator'patterns:addressingarchitecturaluncertaintyininteractivesystems, ATwo-StepApproachforModellingFlexibilityinSoftwareProcesses}.
Additionally, the set of valid combinations of these parts must also be defined, which can possibly increase or reduce the range of valid alternatives.
As a result, the designer is provided with a compact but expressive representation of all solution candidates.
Furthermore, modelling with uncertainty can mean a probabilistic extension of a normal model made by adding probabilities to every assumed value, which are mostly attribute values~\cite{Acamerauncertaintymodelforcollaborativevisualsensornetworkapplications,ExpressingMeasurementUncertaintyinSoftwareModels, AGoal-BasedModelingApproachtoDevelopRequirementsofanAdaptiveSystemwithEnvironmentalUncertainty.,Addinguncertaintyandunitstoquantitytypesinsoftwaremodels, AnApproachforDecisionSupportontheUncertaintyinFeatureModelEvolution}. 
These probabilities are often referred to as confidence values.

Uncertainty can be used in different phases of the modelling process, and there are multiple strategies to eventually resolve uncertainty~\cite{Flexible, MAV-Vis:anotationformodeluncertainty, OnlineModel-BasedTestingunderUncertainty., The'domainmodelconcealer'and'applicationmoderator'patterns:addressingarchitecturaluncertaintyininteractivesystems, TowardsModellingandReasoningAboutUncertainDataofSensorMeasurementsforDecisionSupportinSmartSpaces., TransformationofModelsContainingUncertainty., Softwareengineeringinanuncertainworld, TowardsInverseUncertaintyQuantificationinSoftwareDevelopmentShortPaper, UncertaintymanagementinsoftwareengineeringPastpresentandfuture}.
The lack of information about the content of models is denoted as design-time uncertainty, which makes it impossible to select among alternative design decisions.
This uncertainty can be captured in partial models consisting of a ``base model'' enriched with annotations that express the set of alternatives~\cite{Managingrequirementsuncertaintywithpartialmodels, Languageindependentrefinementusingpartialmodeling, Thesemanticsofpartialmodeltransformations}.
By refining the partial model, uncertainties can be resolved during the design phase~\cite{TowardsaMethodologyforVerifyingPartialModelRefinements}.
The residual uncertainty is denoted as run-time uncertainty and is resolved by the user via a selection out of all remaining alternatives.

Different sources of uncertainty can be distinguished regarding multiple dimensions~\cite{TowardsInverseUncertaintyQuantificationinSoftwareDevelopmentShortPaper, AutomaticallyGeneratingBehavioralModelsofAdaptiveSystemstoAddressUncertainty., Managingnon-functionaluncertaintyviamodel-drivenadaptivity, Relaxingclaims:copingwithuncertaintywhileevaluatingassumptionsatruntime, TowardsUsingProbabilisticModelstoDesignSoftwareSystemswithInherentUncertainty, UncertaintyinSelfAdaptiveSoftwareSystems, UncertaintyWiseCyberPhysicalSystemtestmodeling, Uncertaintywiseevolutionoftestreadymodels}.
The source of uncertainty can either be the system itself or its execution environment.
System uncertainty includes uncertainty about input parameters, structural and algorithmic uncertainty due to approximations, or experimental uncertainty caused by variable measured values.  
Environmental uncertainty can originate from incomplete information about the behaviour of external components, which are provided by third-party organisations, or input data provided by sensors or wireless networks.
Furthermore, the root cause of uncertainty can either be the lack of knowledge about one of the aforementioned factors or some non-determinism within the system.
In general, as uncertainty forces the developer to make decisions based on assumptions, one is not able to guarantee the optimality of those decisions, involving various trade-offs.

In summary, modelling with uncertainty denotes a way of efficiently encoding multiple alternatives into a single representation, from which at least one valid, i.e. consistent, configuration should be derivable.
Tolerance, in contrast, means being able to perform operations on models in the presence of inconsistencies, while the ultimate goal is still to eventually reach a consistent state.
Both concepts aim at facilitating the work flow of system designers and developers by aligning the principles of \ac{MDE} to practical requirements.
Likewise, both tolerance and uncertainty involve a combination of automated and user-centric resolution strategies to finally obtain an unambiguous and consistent model.

	\section{Examples and Application Domains}
\label{sec:examples}

This section gives an overview of examples related to tolerance and uncertainty in different application fields.
Although the underlying notion for consistency is essential for understanding the proposed approaches to tolerating inconsistencies (cf. Sect.~\ref{sec:definitions}), the presented examples and mentioned (dis-)advantages mostly refer to concepts of tolerance and uncertainty building up on it.
Instead, we also found examples and arguments for \textit{flexibility} as a general term subsuming tolerance and uncertainty, such that we devote a further subsection to it in the following.

\begin{table}[htb]
	\begin{footnotesize}
		\begin{tabular}{|c|c|c|c|}
			\hline
			Research Domain & \# & Research Domain & \# \\
			\hline
			Aspect-oriented modelling	&	3	&	Process Modelling				&	4	\\
			(Meta-)model Co-Evolution	&	4	&	Product Line Engineering		&	7	\\
			Cyber-Physical Systems		&	7	&	Requirements Engineering		&	9	\\
			Databases					&	5	&	Service-Oriented Computing		&	3	\\
			Distributed Systems			&	4	&	Smart \& Adaptive Systems		&	10	\\
			Language Engineering		&	4	&	Software Architecture			&	4	\\
			Mobile \& Cloud Computing	&	3	&	Software Engineering (Other)	&	5	\\
			Model-Based Testing			&	7	&	Software Verification			&	4	\\
			Model-Driven Engineering	&	74	&	\textbf{TOTAL}					&	\textbf{157}	\\
			\hline
		\end{tabular}
	\end{footnotesize}
	\caption{Number of relevant papers per research domain}
	\label{tab:relevant-papers-per-research-domain}
\end{table}

In Tab.~\ref{tab:relevant-papers-per-research-domain}, the number of relevant sources per research domain is depicted.
Most of the sources are related to \ac{MDE} and similar fields, such as co-evolution, model-based testing, process modelling, and aspect-oriented modelling.
Especially uncertainty appears to play an important role for requirements engineering and adaptive systems.
The relatively large number of papers concerning other sub-domains of software engineering such as language engineering, product line engineering, software architecture and service-oriented computing underpins the importance of tolerance and uncertainty for the entire field of research.
Cyber-physical and distributed systems, as well as mobile computing, can be identified as relevant application domains due to the substantial impact of environmental conditions involved.
Several papers concern multiple research domains, such that a prioritisation was necessary in these cases:
When \ac{MDE} concepts were applied to a concrete use case, the paper was allocated to the application domain.
For papers which can be matched to different software engineering domains, the main focus was taken as decisive factor.
 
From the set of relevant sources 36 examples could be extracted, which can be used to illustrate approaches to tolerance, uncertainty, flexibility, or consistency management in general.
23 examples focus on a conceptual approach, 6 are used for tool demonstrations, and 7 cover both purposes equally.
In the following, examples for tolerance, uncertainty, and flexibility are briefly sketched; a complete list including examples for consistency management and further classifications can be found online\footnote{\url{https://drive.google.com/file/d/1uSuOn3hX5BHpLhw3jaH2ZpVffgTxHOov}}.

\subsection{Tolerance}
A frequently used example for tolerance is a simplified video-on-demand system modelled with \ac{UML} diagrams~\cite{FineTuningModelTransformationChangePropagationinContextofConsistencyCompletenessandHumanGuidance.,FixingInconsistenciesinUMLDesignModels.,FromAbstracttoConcreteRepairsofModelInconsistencies-AnAutomatedApproach.,GeneratingandEvaluatingChoicesforFixingInconsistenciesinUMLDesignModels.,IncrementalConsistencyCheckingforComplexDesignRulesandLargerModelChanges.,UML/Analyzer-AToolfortheInstantConsistencyCheckingofUMLModels.,AutomaticallyDetectingandTrackingInconsistenciesinSoftwareDesignModels,InstantconsistencycheckingfortheUML,Supportingautomaticmodelinconsistencyfixing, Detectingandexploringsideeffectswhenrepairingmodelinconsistencies}.
The system consists of a streamer retrieving and decoding the content, and a display showing the video and receiving user input.
Each component is modelled with a state chart diagram, in addition to a common class and sequence diagram for both components.
Design rules describe the semantic interrelations between state charts, class diagram, and sequence diagram, e.g. that a class method name be equal to the corresponding message name in the sequence diagram, or that a message sequence match the behaviour in the state chart. 
As a software tool cannot decide if the effects of fixing inconsistencies, probably introduced when changing one model, are desirable or not, a tolerant treatment is suggested.

Tolerating inconsistencies has been a long-term research topic for databases, which is illustrated by an example dealing with a project management tool storing information about the utilisation of employees for projects, as well as how many hours per week they should work~\cite{ToleratingInconsistency.}.
Constraints ensure that the sum of hours an employee works in all projects is equal to their regular working hours.
The time an employee is needed for a project is maintained by project managers, whereas the regular working hours can only be changed by the business office.
It is, therefore, not possible to change project plans or working hours without temporarily introducing inconsistencies.

In the requirements engineering domain, requirements can be described with model fragments that conform to a core requirements metamodel~\cite{ComposingModelsforDetectingInconsistencies-ARequirementsEngineeringPerspective.}.
The complete specification for a library management system, in which books must be registered such that customers can borrow them, can be created by fusing all fragments into one model.
As this procedure can clearly lead to inconsistencies and the loss of meta-model compliance, it is necessary to temporarily relax the metamodel regarding abstract classes, multiplicities, and containment relations.
Metamodel conformance is later restored by fixing the remaining inconsistencies.

The DOPLER tool suite~\cite{ApplyingaConsistencyCheckingFrameworkforHeterogeneousModelsandArtifactsinIndustrialProductLines.} is used in software product line engineering for sales support systems for product configuration.
Based on Eclipse, the tool is able to manage consistency between a variability model, a calculation model, and document templates.
All models can be edited in parallel via suitable editor windows.
Inconsistencies are tolerated in a way that their immediate resolution is not enforced, but occurring problems are listed in the Eclipse error viewer.

Adaptive systems have to cope with the impact of environmental conditions; this is also reflected in their software models.
For flood warning systems, it is important to predict floods as early as possible to reduce damage~\cite{AutomaticallyGeneratingBehavioralModelsofAdaptiveSystemstoAddressUncertainty.}.
In a distributed system of sensors, water depth is calculated with pressure sensors, while flow speed is determined with camera sensors.
The sensor nodes transmit the information to a gateway node, which forwards the predictions to an off-site server.
The system needs to be fault-tolerant because signals can get lost, nodes can get disconnected, etc.
Uncertainty is also involved regarding the execution environment and an appropriate trade-off between functional behaviour (e.g. prediction accuracy) and non-functional characteristics (e.g. energy efficiency) for changing environmental conditions.

\subsection{Uncertainty}
Another use case for systems that dynamically adapt to uncertain environmental conditions is a smart phone app for shop reviews, which provides users with information about lower prices for a product~\cite{Managingnon-functionaluncertaintyviamodel-drivenadaptivity}.
The product's bar code is scanned with the camera to identify the product, and the user's position is determined to make suggestions for nearby shops, while an online search in web shops is performed simultaneously.
However, the quality of the photo, the positioning system and the availability of mobile data represent sources of uncertainty that have to be taken into account when modelling the system.

Uncertainty in model-based testing is demonstrated by testing \ac{UML} specifications for a video conference system~\cite{UncoveringUnknownSystemBehaviorsinUncertainNetworkswithModelandSearch-BasedTesting.}.
The models store information about the number of connected participants and the video quality.
Changing environmental conditions, such as packet loss in the network, or joining and leaving participants, are the primary sources of uncertainty.

The use of type systems can be substantially influenced by the uncertainty of measured values.
In an illustrative case study, a toy car drives along a straight track, which is partitioned into multiple sections.
Within this set-up, the car's velocity and acceleration on each of the sections~\cite{Addinguncertaintyandunitstoquantitytypesinsoftwaremodels} should be computed.
The system model involves uncertainty regarding the length of the sections (at design-time) and the time measurements (at run-time).
Besides these absolute values, also relative values, such as the velocity and acceleration of the car, are uncertain.

Several small-scale but useful examples for modelling with uncertainty originate from \ac{MDE} research itself.
In an e-commerce application for selling books, data about books, authors, comments on the books, and details on books and authors is  shown to the user~\cite{EnhancingFlexibilityinUserInteractionModelingbyAddingDesignUncertaintytoIFML.}.
A user interaction model specifies how a user can navigate between the respective views with help of the \ac{IFML}.
Due to a combinatorial explosion, it is challenging to evaluate all possible alternative flows with usability tests.
Integrating uncertainty in \ac{IFML} models, however, can help to specify a compact encoding of all these alternatives.

Code refactorings can be expressed in software models in terms of transformation rules.
This becomes especially challenging for models incorporating uncertainty~\cite{Thesemanticsofpartialmodeltransformations}.
To explain transformation semantics on uncertain models, it is assumed that a modeller might not be sure whether an attribute should be added to the subclass or the superclass of an inheritance hierarchy.
Furthermore, the model is to be refactored by adding get- and set-methods to both classes, which leads to an exponential growth of possible results, demonstrating the need for a compact encoding of uncertain values.

In a fictional automotive design project, modelling with uncertainty is motivated for \ac{UML} class diagrams.
The three involved classes represent controllers for engine, body, and security of the car~\cite{TowardsaMethodologyforVerifyingPartialModelRefinements}.
Similarly, a perception system for autonomous driving is used to demonstrate uncertainty occurring during object detection and position determination in a scene~\cite{TowardsUsingProbabilisticModelstoDesignSoftwareSystemswithInherentUncertainty}.
In a partial model (involving design uncertainty), each controller's attributes are modelled as attribute sets, which can be refined to discrete attributes by partial model refinement.
Besides attributes, this refinement step can affect the existence of an inheritance relation (e.g. between the classes car and vehicle) or the knowledge if car and vehicle are actually the same class~\cite{Languageindependentrefinementusingpartialmodeling}.

Smart home systems provide solutions for intrusion detection with sensors, which are however exposed to uncertainty on several levels. 
Both imprecise measurements on sensors, the network infrastructure which connects the sensors and the interactions between physical units can be sources of uncertainty, leading to false positives and negatives when triggering alarm signals~ 
\cite{Model-BasedTestingUnderParametricVariabilityofUncertainBeliefs, UncertaintyWiseCyberPhysicalSystemtestmodeling}.


The design of an automatic reasoning engine for logical expressions is taken as an example for design uncertainty~\cite{TransformationofModelsContainingUncertainty.}.
When the reasoning engine reaches an error state, a solver exception should be thrown, whose concrete implementation has some points of uncertainty.
The exception may be an inner class of the solver, or an attribute could possibly provide more information about the error type.
A similar example for uncertainty resulting from incomplete requirements is presented via an \ac{UML} state chart for a bank ATM. 
Depending on the required level of strictness, the ATM can either be restarted or set to be out of service in case of errors~\cite{Uncertaintyinbidirectionaltransformations}.

Finally, a framework for model-based testing under uncertain conditions was presented in recent work~\cite{UncertaintyWiseCyberPhysicalSystemtestmodeling} to cope with the inherent uncertainty of \ac{CPS} components. 
Connecting it to a test ready model evolution framework~\cite{Uncertaintywiseevolutionoftestreadymodels}, it is possible to generate further test cases for \acp{CPS} from evolved models.

\subsection{Flexibility}
Some of the considered examples illustrate the use of flexibility in software modelling, which can be seen as a generalisation of concepts including both tolerance and uncertainty.
Motivating examples for flexibility with respect to metamodel conformance are proposed in co-evolution scenarios.
When keeping class diagrams and relational databases consistent~\cite{Consistentco-evolutionofmodelsandtransformations.}, refactorings on the metamodel make it necessary to adapt the transformation definition and the models, which should comply to this modified metamodel.
As common examples for refactorings, deleting or moving attributes to other classes, introducing inheritance relations, or renaming references are listed.

In a similar setting, a family register is to be kept consistent with a persons register~\cite{JSMFaJavascriptFlexibleModellingFramework}, such that, for example, the first and last name of a family member should be consistent with the full name of the corresponding person.
When the metamodel is adapted, e.g. by adding a nickname attribute to the family member or by fusing first and last name, it is useful to relax the conformance relation by (temporarily) deactivating type or cardinality checks.

Especially when working with EMF, co-evolving metamodels introduce problems and additional effort for the persons involved, which is illustrated by a small case study modelling the network infrastructure of an office, including all shared gadgets such as scanners, photocopiers and fax machines~\cite{EnhancedAutomationforManagingModelandMetamodelInconsistency.}.
As the EMF editor always enforces strict metamodel conformance, it is not possible to work with models conforming to older versions of the metamodel.
A common workaround is the trial-and-error strategy of loading a model to get an error message from EMF, and then attempting to fix this error directly in the XMI document, obviously a tedious and error-prone task.

To model flexibility in software processes, the \ac{EPFC} was extended to ease the collaboration of the involved persons~\cite{ATwo-StepApproachforModellingFlexibilityinSoftwareProcesses}.
The process engineers propose a flexible work flow, which can be adapted by other participants in a second step.
%

	\section{Benefits and Challenges}
\label{sec:benefits-drawbacks}

To investigate the third research question, arguments were collected that support or question the use of tolerance or uncertainty.
As benefits and drawbacks differ, the two concepts were analysed separately.
More general arguments, which deal with more flexibility in software engineering, concern both concepts and are discussed afterwards.

\subsection{Tolerance}
A range of benefits resulting from tolerating inconsistencies to some extent was identified during the \ac{SLR}.
In some application scenarios, such as distributed software systems, fault-tolerance is required to achieve availability and partition-tolerance~\cite{AConsistencyModelforIdentityInformationinDistributedSystems., AFormalModelforFault-ToleranceinDistributedSystems., DataQualityMaintenancebyIntegrity-PreservingRepairsthatTolerateInconsistency., ToleratingInconsistency.}.
Similarly, being able to handle inconsistencies is essential due to the modularity of applications and data sources in modern software systems; errors can easily occur when composing building blocks in a new way, even if each module is implemented correctly~\cite{AMethodforModelingandAnalyzingFault-TolerantServiceComposition., Classifyingintegritycheckingmethodswithregardtoinconsistencytolerance.}.
A frequently mentioned point is that temporarily tolerating inconsistencies can ease the work flow for system designers and testers as a fault-tolerant \ac{IDE} does not enforce the restoration of consistency before further modelling steps can be performed~\cite{ApplyingaConsistencyCheckingFrameworkforHeterogeneousModelsandArtifactsinIndustrialProductLines., Automatedconsistencyandcompletenesscheckingoftestingmodelsforinteractivesystems, ComposingModelsforDetectingInconsistencies-ARequirementsEngineeringPerspective., EMA2AOP-FromtheAADLErrorModelAnnextoaspectlanguagetowardsfaulttolerantsystems., OntheQuestforFlexibleModelling., RuntimeAdjustmentofConfigurationModelsforConsistencyPreservation., AutomaticallyDetectingandTrackingInconsistenciesinSoftwareDesignModels, CAREaconstraintbasedapproachforreestablishingconformancerelationships}.
Usually, atomic changes such as graph edits can lead to inconsistent states in between, which should be tolerated at least until the entire edit sequence is completed~\cite{Anautomatedandinstantdiscoveryofconcreterepairsformodelinconsistencies., BidirectionallyToleratingInconsistency-PartialTransformations., DataQualityMaintenancebyIntegrity-PreservingRepairsthatTolerateInconsistency., Designandevaluationofacontinuousconsistencymodelforreplicatedservices, EMA2AOP-FromtheAADLErrorModelAnnextoaspectlanguagetowardsfaulttolerantsystems., Flexibleandscalableconsistencycheckingonproductlinevariabilitymodels., IncrementalConsistencyCheckingforComplexDesignRulesandLargerModelChanges., Modelingexecutiontimeofmulti-stageN-versionfault-tolerantsoftware., OntheQuestforFlexibleModelling., UsingCritiquingSystemsforInconsistencyDetectioninSoftwareEngineeringModels., AutomaticallyDetectingandTrackingInconsistenciesinSoftwareDesignModels, Makinginconsistencyrespectableinsoftwaredevelopment, Detectingandexploringsideeffectswhenrepairingmodelinconsistencies}.
From a practical point of view, improving consistency might be more helpful than enforcing it strictly.
Often, the restoration process requires multiple changes that can each be regarded as an improving step towards consistency, while only the last one is able to finally restore consistency~\cite{BidirectionallyToleratingInconsistency-PartialTransformations., DataQualityMaintenancebyIntegrity-PreservingRepairsthatTolerateInconsistency.}.
Furthermore, a detected error often reveals another problem, which might be the root cause of multiple other defects.
Therefore, information about inconsistencies is often more helpful than an automated fix that achieves consistency~\cite{GeneratingandEvaluatingChoicesforFixingInconsistenciesinUMLDesignModels., UML/Analyzer-AToolfortheInstantConsistencyCheckingofUMLModels.,Detectingandexploringsideeffectswhenrepairingmodelinconsistencies}.
In fault-tolerant systems, the number of automated changes can be decreased, which can improve the tool's trustworthiness for the designer~\cite{BidirectionallyToleratingInconsistency-PartialTransformations.}.
Even if automated changes are the preferred way of resolving conflicts, their application often relies on a central authority to define a policy for restoration steps.
Especially when more than two models are involved, it is in general not possible to declare one of the models as the authoritative one, or prefer a certain type of changes over others due to, e.g., transitive consequences~\cite{Declarativeprojectplanningandcontrolling-aformalmodeltosupportthehandlingofunavoidableinconsistencies., Towardssoundoptimalandflexiblebuildingfrommegamodels}.
It follows that, whenever a design decision is ambiguous, only the uncontroversial steps can be fully automated - leading to a possibly inconsistent state - before the user must participate in resolving the remaining inconsistencies~\cite{BidirectionallyToleratingInconsistency-PartialTransformations., FineTuningModelTransformationChangePropagationinContextofConsistencyCompletenessandHumanGuidance., Supportingautomaticmodelinconsistencyfixing}.
Another set of arguments refers to weaknesses of fixing-procedures.
To maintain an acceptable level of efficiency, many approaches apply local fixes to restore consistency.
This means that mechanisms don't have a global view on the modelled system, and local fixes can have side-effects that are not monitored by the tool~\cite{FineTuningModelTransformationChangePropagationinContextofConsistencyCompletenessandHumanGuidance., FixingInconsistenciesinUMLDesignModels.,QuickfixgenerationforDSMLs}.
Consequently, these fixes might introduce new inconsistencies, which are often hard to detect, and which have to be fixed at a later point~\cite{DataQualityMaintenancebyIntegrity-PreservingRepairsthatTolerateInconsistency., FixingInconsistenciesinUMLDesignModels., GeneratingandEvaluatingChoicesforFixingInconsistenciesinUMLDesignModels., ImprovingInconsistencyResolutionwithSide-EffectEvaluationandCosts., AutomaticallyDetectingandTrackingInconsistenciesinSoftwareDesignModels,Detectingandexploringsideeffectswhenrepairingmodelinconsistencies, QuickfixgenerationforDSMLs}.
As multiple stakeholders are involved in the modelling of complex systems, their requirements can be contradictory.
Without being able to tolerate these defects for a while, the modelling process gets stuck at this point and requires an instant resolution~\cite{Automaticdetectionofincompleterequirementsviasymbolicanalysis, InstantconsistencycheckingfortheUML}.
Last but not least, the consistency check itself can be erroneous, such that the respective tool finds false positives.
While it is undisputed that the error has to be removed, tolerant behaviour could again ease the continuation of the modelling process~\cite{Inconsistencymanagementframeworkformodel-baseddevelopment.}.

Despite this long list of advantages, many authors argue against involving tolerance in system design.
It is questionable how long and to which extent inconsistencies should be tolerated, because a large number of factors have an influence on the value of fault-tolerance in a specific use case~\cite{AConsistencyModelforIdentityInformationinDistributedSystems., ComposingModelsforDetectingInconsistencies-ARequirementsEngineeringPerspective., Detectingmodelinconsistencythroughoperation-basedmodelconstruction., FixingInconsistenciesinUMLDesignModels., FromAbstracttoConcreteRepairsofModelInconsistencies-AnAutomatedApproach., UseCase-BasedModelingandAnalysisofFailsafeFault-Tolerance.}.
Certainly, one should not lose track of the goal of eventually restoring consistency.
Assuming that it is always possible to fix inconsistencies at a later point, this might still involve additional effort and thus a higher cost~\cite{UML/Analyzer-AToolfortheInstantConsistencyCheckingofUMLModels., InstantconsistencycheckingfortheUML, Computingrepairtreesforresolvinginconsistenciesindesignmodels., Managingrequirementsuncertaintywithpartialmodels, PrototypinganInconsistencyCheckingToolforSoftwareProcessModels., Detectingandexploringsideeffectswhenrepairingmodelinconsistencies}.
On the one hand, when consistency should eventually be restored, the developer might be confronted with so many errors at once that they are overwhelmed.
On the other hand, errors might be caused by changes that occurred so long ago, that design decisions have to be completely revisited~\cite{InstantconsistencycheckingfortheUML}.
As developers are usually willing to fix errors as soon as they are noticed, they will probably do the same when they detect inconsistencies that could actually be tolerated by the system.
This means that such a tool's potential for supporting tolerance will probably be ignored by its users~\cite{DetectingandRepairingInconsistenciesacrossHeterogeneousModels.}.
Also, performing operations (e.g. model transformations) on inconsistent models likely introduces further errors~\cite{Detectingandexploringsideeffectswhenrepairingmodelinconsistencies}.
Finally, many tools for model management are based on formal methods, which are not yet compatible with tolerant concepts~\cite{Classifyingintegritycheckingmethodswithregardtoinconsistencytolerance.}.

\subsection{Uncertainty}
An important argument for modelling with uncertainty is that it represents real-world scenarios more accurately.
The input data and the behaviour of \acp{CPS} and adaptive systems is imprecise, e.g. their sensors and actuators provide the system with imprecise data, and this should be reflected in a model of the system~\cite{Acamerauncertaintymodelforcollaborativevisualsensornetworkapplications, Addinguncertaintyandunitstoquantitytypesinsoftwaremodels, AutomaticallyGeneratingBehavioralModelsofAdaptiveSystemstoAddressUncertainty., ExpressingMeasurementUncertaintyinSoftwareModels, OnlineModel-BasedTestingunderUncertainty., TestModelCoverageAnalysisUnderUncertainty., TowardsModellingandReasoningAboutUncertainDataofSensorMeasurementsforDecisionSupportinSmartSpaces., U-Test:Evolving, TowardsInverseUncertaintyQuantificationinSoftwareDevelopmentShortPaper, UncertaintyinSelfAdaptiveSoftwareSystems}.
Similarly, information might not be available in distributed systems, or not accessible due to security restrictions or authentication problems~\cite{DataUncertaintyModelforMashup.}.
In software and requirements engineering processes, uncertainty plays an important role, especially in early phases.
When designing complex and widely heterogeneous systems, multiple stakeholders are involved, whose understanding of the final result may be incomplete~\cite{AGeneralizedSoftwareReliabilityModelConsideringUncertaintyandDynamicsinDevelopment, Flexible, Mu-Mmint:anIDEformodeluncertainty, MAV-Vis:anotationformodeluncertainty, OnlineModel-BasedTestingunderUncertainty., Partialmodels-Towardsmodelingandreasoningwithuncertainty., Relaxingclaims:copingwithuncertaintywhileevaluatingassumptionsatruntime, TechneTowardsaNewGenerationofRequirementsModelingLanguageswithGoalsPreferencesandInconsistencyHandling, The'domainmodelconcealer'and'applicationmoderator'patterns:addressingarchitecturaluncertaintyininteractivesystems, TransformationofModelsContainingUncertainty., UsingModelsatRunTimetoDetectIncompleteandInconsistentRequirements., Softwareengineeringinanuncertainworld,Model-BasedTestingUnderParametricVariabilityofUncertainBeliefs,Uncertaintyinbidirectionaltransformations}.
Additionally, this incomplete information makes it necessary to continuously adapt the development process and the requirements~\cite{AGoal-BasedModelingApproachtoDevelopRequirementsofanAdaptiveSystemwithEnvironmentalUncertainty., DetectingandResolvingModelInconsistenciesUsingTransformationDependencyAnalysis., Managingrequirementsuncertaintywithpartialmodels, OnlineModel-BasedTestingunderUncertainty., Relaxingclaims:copingwithuncertaintywhileevaluatingassumptionsatruntime, Softwareengineeringinanuncertainworld, Thesemanticsofpartialmodeltransformations, TowardsaMethodologyforVerifyingPartialModelRefinements}.
Likewise, removing uncertainty too early can force the designer to commit to premature decisions that can increase cost and efforts to remove resulting errors later~\cite{TransformationofModelsContainingUncertainty.,Uncertaintyinbidirectionaltransformations},
while ignoring uncertainty completely decreases the overall quality of the software~\cite{UncertaintymanagementinsoftwareengineeringPastpresentandfuture, UncertaintyinSelfAdaptiveSoftwareSystems}.
An obvious alternative to model uncertainty is to list each alternative value explicitly, but this can quickly become infeasible.
Uncertainty is an elegant way to encode alternatives and non-determinism, while still keeping models manageable~\cite{EnhancingFlexibilityinUserInteractionModelingbyAddingDesignUncertaintytoIFML., HandlingUncertaintyinAutomaticallyGeneratedImplementationModelsintheAutomotiveDomain, Managinguncertaintyinbidirectionalmodeltransformations, DeterminingtheCauseofaDesignModelInconsistency}.
Indeed, uncertain values are probably easier to maintain than a large set of alternatives~\cite{HandlingUncertaintyinAutomaticallyGeneratedImplementationModelsintheAutomotiveDomain, Managingnon-functionaluncertaintyviamodel-drivenadaptivity}.
In case some design choices are more likely to be applied than others, uncertainty is also useful to express these varying probabilities quantitatively~\cite{AnEmpiricalApproachtoModelingUncertaintyinIntrusionAnalysis}. 
From a practical point of view, uncertainty is often indirectly added to models via informal annotations in case its direct expression is not supported.
Therefore, enabling the designer to model uncertainty in the given formal notation improves the verification and validation of such models~\cite{TowardsaMethodologyforVerifyingPartialModelRefinements}.

Handling uncertainty, however, can also increase development cost as the encoded set of alternatives - which takes all possible combinations of values into account - can be much larger than the set of possible options in practice~\cite{AGoal-BasedModelingApproachtoDevelopRequirementsofanAdaptiveSystemwithEnvironmentalUncertainty., The'domainmodelconcealer'and'applicationmoderator'patterns:addressingarchitecturaluncertaintyininteractivesystems,UncertaintymanagementinsoftwareengineeringPastpresentandfuture}.
Following the same argument, the model space grows exponentially with the degree of uncertainty, which can cause performance problems for larger model sizes~\cite{AnEmpiricalApproachtoModelingUncertaintyinIntrusionAnalysis, Partialmodels-Towardsmodelingandreasoningwithuncertainty.,UncertaintyinSelfAdaptiveSoftwareSystems}.
Even though uncertainty is an appropriate way of expressing probabilities, it can be difficult to realistically quantify uncertainty measures, as empirical tests for these measures are often missing or cannot be conducted at all~\cite{AnEmpiricalApproachtoModelingUncertaintyinIntrusionAnalysis,UncertaintyinSelfAdaptiveSoftwareSystems}.
Finally, operations on models are usually designed for single models, whereas uncertain models encode a whole set, restricting the applicability of standard tooling~\cite{Partialmodels-Towardsmodelingandreasoningwithuncertainty.}.

\subsection{Flexibility}
It is possible that multiple consistent solutions exist, which deviate in their quality.
As it is hard to specify which solution should be taken, the system should provide the flexibility to let the user make this final decision~\cite{ADynamic-PriorityBasedApproachtoFixingInconsistentFeatureModels.}.
To keep the complexity of a system manageable, software is usually developed with an idealised environment in mind.
However, the system is also expected to react robustly to environment changes and unforeseen circumstances at runtime~\cite{AModelingApproachforFlexibleWorkflowApplicationsofCloudServices., AnIncrementalAlgorithmforHigh-PerformanceRuntimeModelConsistency., The'domainmodelconcealer'and'applicationmoderator'patterns:addressingarchitecturaluncertaintyininteractivesystems, TowardsModel-DrivenVariability-BasedFlexibleServiceCompositions.}.
In application domains such as product line engineering, ``hard constraints'' can reduce the potential of the modelling language and therefore restrict the scope of action for the designer~\cite{FlexibleModelingandProductDerivationinSoftwareProductLines., Addingflexibilityinthemodel-drivenengineeringofuserinterfaces}.
In the area of model-metamodel co-evolution, some flexibility is necessary for a modelling tool to be appropriate for practical use.
The temporary loss of metamodel conformance should not lead to a situation where the model cannot be modified or even loaded in the respective editor~\cite{EnhancedAutomationforManagingModelandMetamodelInconsistency., TheConformanceRelationChallenge-BuildingFlexibleModellingFrameworks., AMetamodelingFrameworkforPromotingFlexibilityandCreativityOverStrictModelConformance, AmultilevelapproachtomodelinglanguageextensionintheEnterpriseSystemsDomain, ApproachestoCoEvolutionofMetamodelsandModelsASurvey, Constraintprogrammingfortypeinferenceinflexiblemodeldrivenengineering, JSMFaJavascriptFlexibleModellingFramework}.
Finally, the result of a model transformation is often not unique, requiring a flexible encoding~\cite{Asatisficingbidirectionalmodeltransformationengineusingmixedintegerlinearprogramming, Uncertaintyinbidirectionaltransformations}.

A few arguments can also be found that question the benefits of flexibility.
As tools are typically not built by the intended users, the developer might have a different understanding of flexibility, such that the user may ignore or even disregard any support for it~\cite{Amodel-drivenapproachtoflexiblemulti-levelcustomizationofSaaSapplications.}.
The more flexibility is added to a system, the more complexity is involved as well, which can end up in a misinterpretation of functionality or a loss of overview while developing and maintaining the tool~\cite{ConsistencyCheckinModellingMulti-AgentSystems., EvaluatingtheImpactofAspectsonInconsistencyDetectionEffort-AControlledExperiment., MAV-Vis:anotationformodeluncertainty}
Finally, in case of errors and other inconsistencies, software developers are currently used to instant feedback from \acp{IDE} for general purpose languages, and will probably expect similar behaviour from modelling tools.
Transitioning to tolerant tooling will therefore require a certain retraining of users to ensure acceptance, and it is still unclear how challenging this will be in practice~\cite{InstantconsistencycheckingfortheUML}.
	\section{Result Analysis}
\label{sec:result-analysis}

This section provides an overview of aggregated meta-data of the considered sources, before directions for future research are sketched, which can be motivated by this \ac{SLR}. 

\begin{figure}[htb]
	\includegraphics[width=\columnwidth]{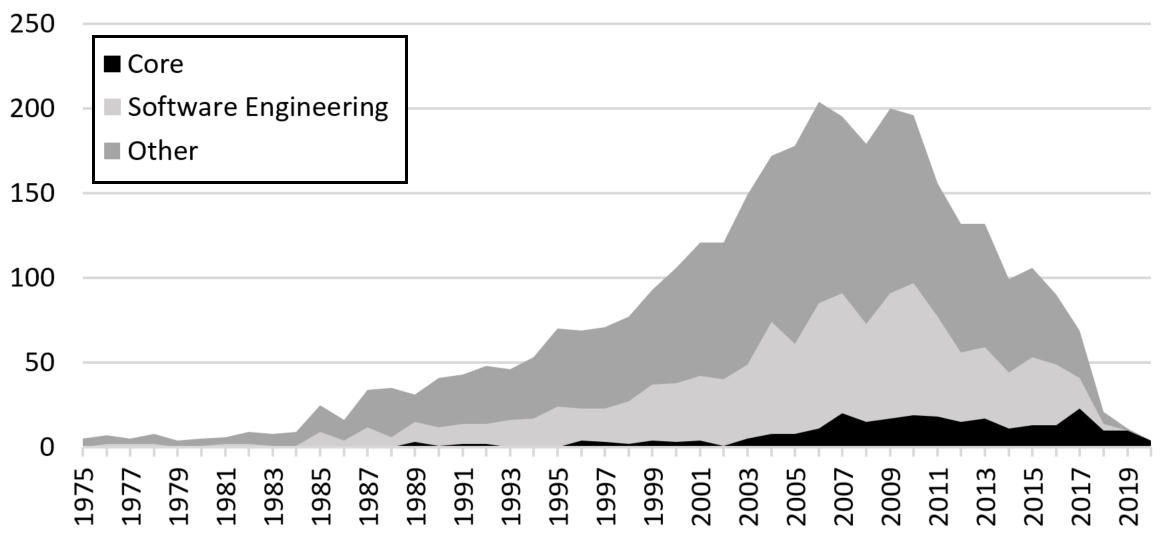}
	\caption{Number of sources per year}
	\label{fig:all-papers}
\end{figure}

The distribution of all sources in the database, i.e. all core papers and all sources cited by them, is depicted in Fig.~\ref{fig:all-papers}, whereby ten sources published before 1975 are not captured in the diagram.
In total, 268 core papers, 1146 other papers at \ac{SE} venues and 2055 other sources were found in the initial search step.
The median (mean) publication year is 2010.5 (2009.78) for core papers, 2007 (2005.87) for other papers at \ac{SE} venues and 2006 (2004.35) for the remaining sources.
It follows that filtering the additional sources was necessary to keep the amount of work manageable, and also that being published at a venue listed as research area 0803 (software engineering) is a useful indicator for increased relevance; at least one third of the additional sources was published at such a venue.
The differences between core papers and other sources with respect to the average publication year can be explained by the applied snowballing technique, by which only sources released prior to the initial source can be found.
Furthermore, the set of sources published at non-\ac{SE} venues include standard references in form of books and journal articles, which is probably the reason why the papers from \ac{SE} venues are slightly newer on average.

\begin{figure}[htb]
	\includegraphics[width=\columnwidth]{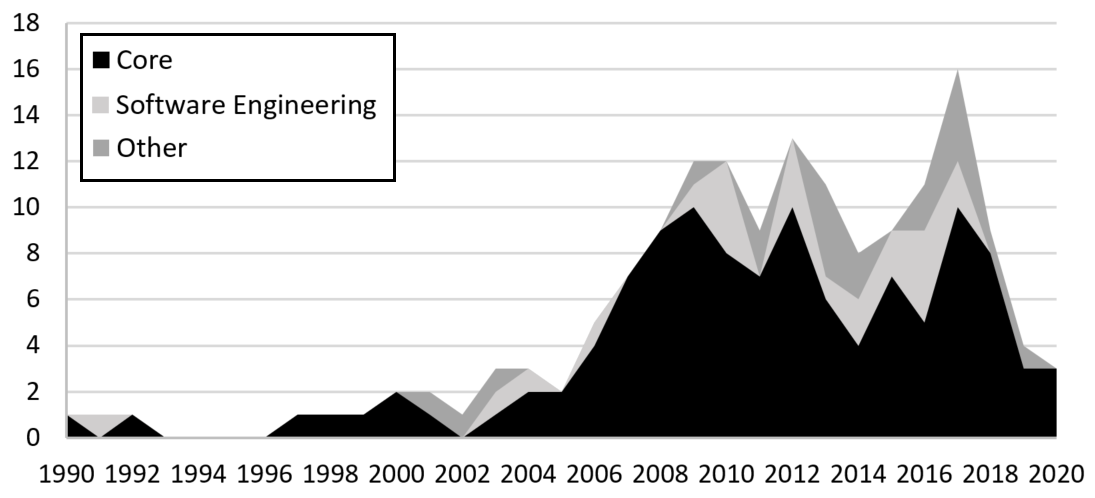}
	\caption{Number of relevant sources per year}
	\label{fig:relevant-papers}
\end{figure}

When taking only those sources into account that were later identified as being relevant for answering the research questions, the majority of sources originates from the set of core papers (cf. Fig.~\ref{fig:relevant-papers}).
Besides 114 of the core papers, 23 papers published at \ac{SE} venues and 20 other sources were classified as relevant.
Compared to the full corpus, the relevant papers are newer on average:
The median (mean) publication year is 2011 (2011.05) for core papers, 2012 (2011.08) for publications at \ac{SE} venues, and 2012 (2011.29) for the remaining relevant sources.
An explanation can be that the search terms are used in a different meaning more frequently in older sources, according to our experience.
Furthermore, as the research field \ac{MDE} became popular along with the emergence of the \ac{UML} in the late 1990s, sources published before can only be relevant for our purposes if they describe transferable concepts or examples from other domains.

Table~\ref{tab:top-conferences} gives an overview of the number of relevant papers per venue, listing those venues with at least four relevant papers.
As this \ac{SLR} deals with a subtopic of \ac{MDE}, finding the MODELS conference at the top of the ranking is a result one would expect.
Five papers published at co-located events were relevant for this \ac{SLR} as well.
ICSE and ASE as two top-ranked \ac{SE} conferences in the list underpin the topic's relevance for a broader audience.
The appearance of important conferences for more specialised research fields, such as requirements engineering (RE), software language design (SLE) and software testing (ICST), shows that fault-tolerance concerns the entire software development process. 
The remaining well-known \ac{SE} venues COMPSAC, SEKE, APSEC and FASE complete the list of venues with 4 or more relevant papers.

\begin{table}[htb]
	\centering
	\begin{scriptsize}
		\begin{tabular}{|c|c|l|c|}
			\hline
			Rank	& Acronym	& Venue	& \# \\
			\hline
			1		&	MODELS	&	Model Driven Engineering: Languages and Systems				&	20 \\
			2		&	ICSE	&	International Conference on Software Engineering			&	15 \\
			3		&	COMPSAC	&	Computer Software and Applications Conference				&	10 \\
			4		&	ASE		&	Automated Software Engineering								&	8 \\
			5		&	SEKE	&	Software Engineering and Knowledge Engineering				&	7 \\
			6		&	FASE	&	Fundamental Approaches to Software Engineering				&	6 \\
			7		&	APSEC	&	Asia-Pacific Software Engineering Conference				&	5 \\
			7		&	RE		&	Intern. Conference on Requirements Engineering 				&	5 \\
			7		&	-		&	MODELS Satellite Events										&	5 \\
			10		&	SLE		&	Software Language Engineering								&	4 \\
			10		&	ICST	&	International Conference on Software Testing				&	4 \\
			\hline
		\end{tabular}
	\end{scriptsize}
	\caption{Top 10 conferences by number of relevant papers}
	\label{tab:top-conferences}
\end{table}


Although as a result of this \ac{SLR}, many thorough definitions, useful examples and convincing arguments for tolerance in \ac{MDE} could be extracted, a need for further research became apparent simultaneously.
As tolerance was mostly defined intuitively, an extended formal framework (e.g. for softening domain constraints) would be helpful to prove properties of tolerant systems.
This includes quantitative measures for (in)consistency, as well as quality criteria for the intermediate and final model states to assess the utility of proposed approaches.
Up to now, it remains also unclear when exactly consistency shall be restored, up to which point errors can be tolerated, and how to deal with conflicting changes that were made in the meantime.
Especially for more than two models, consistency restoration is a non-trivial problem because removing errors in one model can introduce other errors in different places.
Further user studies could show to which extent user interaction is required to resolve such conflicts, and which restoration actions can be performed automatically.
Several authors pointed out that some faults are too serious to be tolerated, but still strategies are needed to assess the severity of errors in a model, though.
While model transformations in presence of uncertainty are already investigated, fault-tolerant consistency management is still an unsolved issue.
Finally, although tolerance and uncertainty could be differentiated in spite of their common goal, it would be useful to specify which of the two concepts is more helpful in which particular scenarios.

	\clearpage

\section{Related Work}
\label{sec:related-work}
Studies that provide a structured overview of existing work on a particular topic are often conducted as \acp{SLR}~\cite{Kitchenham2004} or as mapping studies~\cite{Petersen2008}.
\acp{SLR} are a secondary study that identifies, analyses and interprets all available information related to one or more research  questions.
\acp{SLR} follow a predefined review protocol, such that the process of retrieving results is transparent and the introduced bias is minimised.
Mapping studies categorise existing work, often leading to a visual mapping of categories that supports the understanding of what is already addressed in a specific domain.

Several of these studies of either type have been conducted in the \ac{MDE} domain and related fields.
Modelling languages were investigated with a focus on the \ac{SysML} language~\cite{Wolny2020}, the \ac{QVTo} standard~\cite{Gerpheide2014, Gerpheide2016}, and the application of modelling in Industry 4.0~\cite{Wortmann2020,Wortmann2017}.
Further \ac{MDE}-related work investigates literature on models at runtime~\cite{Szvetits2016}, software testing process models~\cite{Vukovic2018}, articles that appeared in the Journal of Software and Systems Modelling~\cite{Gray2016}, and quality in \ac{MDE}~\cite{Goulao2016}.
In the requirements engineering domain, studies on software tooling for requirement elicitation~\cite{Iqbal2019} and software testing in the context of agile software development~\cite{Coutinho2019} have been presented.
For software product lines, existing work on the automated analysis of feature models~\cite{Galindo2019}, variability management~\cite{Galster2014}, and tool support~\cite{Bashroush2017} has been already investigated.
Context modelling~\cite{Koc2014} and environment modelling~\cite{Siavashi2015} are further topics of existing \acp{SLR}.

Multiple studies on consistency in modelling languages have been already presented.
Awadid et al.~\cite{Awadid2019, Awadid2016} composed an overview of consistency requirements of business process models by proposing a framework for the categorization of approaches and a road-map for future research on consistency requirements elicitation and management.
The work of Muram et al.~\cite{Muram2017} takes consistency checking of software behavioural models into account.
Seven main categories for consistency checking in this domain were identified, and suggestions for future research in this direction were proposed.
Hoisl et al.~\cite{Hoisl2015} conducted a literature review on consistency rules for \ac{UML}-based language models, discussing frequently-named defects of such models and demanding more tool support for enforcing consistency rules in this setting.
All of these studies focus on a sub-domain of \ac{MDE} and do not take tolerance or uncertainty into account.  

Only two studies on existing work relating fault-tolerance to software engineering problems could be found.
Nascimento et al.~\cite{Nascimento2014} analysed literature on the design of fault-tolerant \ac{SOA} using design diversity, deriving guidelines for fault-tolerant \ac{SOA} design and proposing a taxonomy for useful techniques in this respect.
A mapping study for fault-tolerant \ac{IoT} applications~\cite{Moghaddam2019} identifies key factors for tolerant systems, including the use of micro-services and the distribution of \ac{IoT} components.
Both studies are neither directly related to \ac{MDE} nor address the problem of maintaining consistency.

In a study combining an \ac{SLR}, semi-structured interviews, and an empirical evaluation, Marinho et al.~\cite{Marinho2018, Marinho2015} propose and evaluate techniques to distinguish risks and uncertainties to reduce the latter in software projects.
Salih et al.~\cite{Salih2017} provide an overview of existing work on uncertainty involved in requirements engineering via a categorisation of relevant sources, while several questions are left open.
Measurement uncertainty was studied in depth by da Silva Hack et al.~\cite{Hack2012}, resulting in a classification of approaches and a list of methods for calculating uncertainty.
However, these treatments focus solely on uncertainty, whereas tolerance and software modelling are not considered.

As previously mentioned, \acp{SLR} in computer science often lack adequate tool support; this issue has been identified and discussed by existing work.
Götz proposes a tool for processing the findings of \acp{SLR}~\cite{Goetz2018}, which enables the user to assign the relevant papers to formed categories, such that diagrams can be generated that visualize the characteristic values for one or two categories.
The tool supports \acp{SLR} in a later phase, though, as the list of relevant sources is required as input data.
The SLR-Tool by Fernández-Sáez et al.~\cite{Fernandez2010} supports the process of conducting \acp{SLR} in different phases.
Relevant meta-data can be stored for each source, a classification scheme can be created, and diagrams for result visualisation can be exported.
In contrast to our tool-chain, the sources have to be imported manually in the beginning, and support for automated snowballing is not provided.
	\section{Conclusion and Future Work}
\label{sec:conclusion-future-work}

We presented the results of an \ac{SLR} on tolerance in \ac{MDE}, which took 157 relevant sources into account.
The key terms consistency, tolerance, and uncertainty were defined and represented in feature models, such that salient differences and commonalities between tolerance and uncertainty could be pointed out.
Typical use cases for tolerant and uncertain modelling were sketched, and benefits and challenges of the respective concepts were discussed.
To ease the reproducibility of our results and to support future \acp{SLR} in computer science, we proposed a model-driven tool-chain based on open-source components under active development.

Although some relevant journal articles and workshop papers were identified by the snowballing step, the set of sources can be further extended by more sources published at other venues.
Since the CORE2020 journal ranking was recently made available, we plan to extend the literature review towards journal papers following the search strategies presented in this paper.
To the same end, other research databases could be considered as well.
The review has shown that there are indeed useful examples for applying tolerant concepts in \ac{MDE}; a systematic benchmark for comparing tolerant approaches is, however, still an open issue.
	
	\section*{Acknowledgements}
	We would like to thank all anonymous reviewers for their helpful feedback that improved the quality of this \ac{SLR}.
	
	\bibliography{literature}
	
	
\end{document}